\documentclass[
    reprint,
    preprintnumbers,
    superscriptaddress,
    nofootinbib,
     amsmath,amssymb,
     aps,
     prd,
    floatfix,
    ]{revtex4-2}
    
\pdfoutput=1
\usepackage{graphicx,amsmath,amssymb,amstext}
\usepackage{amssymb,amsbsy,amsfonts,amsthm,color}

\usepackage{epsfig}
\usepackage{graphicx}
\usepackage{subfigure}
\usepackage{hyperref}
\usepackage{cancel}
\usepackage{multirow}
\usepackage{parskip}

\graphicspath{{figures/}}

\hypersetup{
  colorlinks   = true, 
  urlcolor     = blue, 
  linkcolor    = blue, 
  citecolor   = blue 
}

\usepackage{color}
\usepackage[dvipsnames]{xcolor}

\newcommand{\mpl}{M_\mathrm{Pl}}
\newcommand{\dd}{\mathrm{d}}

\newcommand{\efolds}{\textit{e}-folds}

\newcommand{\oxford}{Astrophysics, University of Oxford, DWB, Keble Road, Oxford OX1 3RH, United Kingdom}
\newcommand{\qmul}{Geometry, Analysis and Gravitation, School of Mathematical Sciences, Queen Mary University of London, Mile End Road, London E1 4NS, United Kingdom}
\newcommand{\kcl}{Theoretical Particle Physics and Cosmology Group, Physics Department, King's College London, Strand, London WC2R 2LS, United Kingdom}
\newcommand{\ucsd}{UC San Diego, Department of Physics, 9500 Gilman Rd, La Jolla, CA, 92093, USA}
\newcommand{\upv}{Department of Physics, University of Basque Country, UPV/EHU, 48080, Bilbao, Spain}

\begin{document}

\title{Robustness of inflation to kinetic inhomogeneities} 

 \author{Matthew Elley}
 \email{matthew.elley@ehu.eus}
 \affiliation{\upv}
 \affiliation{\kcl}

 \author{Josu C. Aurrekoetxea}
 \email{josu.aurrekoetxea@physics.ox.ac.uk}
 \affiliation{\oxford}
 
 \author{Katy Clough} \email{k.clough@qmul.ac.uk}
 \affiliation{\qmul}
 
 \author{Raphael Flauger}
 \email{flauger@ucsd.edu}
 \affiliation{\ucsd}
 
 \author{Panagiotis Giannadakis}
 \email{panagiotis.giannadakis@kcl.ac.uk}
 \affiliation{\kcl}

 \author{Eugene A. Lim}
 \email{eugene.a.lim@gmail.com}
 \affiliation{\kcl}
\begin{abstract}
We investigate the effects of large inhomogeneities in both the inflaton field and its momentum. We find that in general, large kinetic perturbations reduce the number of \efolds{} of inflation. In particular, we observe that inflationary models with sub-Planckian characteristic scales are not robust even to kinetic energy densities that are sub-dominant to the potential energy density, unless the initial field configuration is sufficiently far from the minimum. This strengthens the results of our previous work. In inflationary models with super-Planckian characteristic scales, despite a reduction in the number of \efolds{}, inflation is robust even when the potential energy density is initially sub-dominant. For the cases we study, the robustness of inflation strongly depends on whether the inflaton field is driven into the reheating phase by the inhomogeneous scalar dynamics.

\end{abstract}
\maketitle

\section{Introduction} \label{sect:intro}

The theory of cosmic inflation \cite{Guth:1980zm, Starobinsky:1980te, Linde:1981mu,Albrecht:1982wi, Linde:1983gd} is most widely studied paradigm of the very early universe. It posits a period of nearly exponential expansion, driving the universe to a state of spatial flatness and homogeneity. Furthermore, it provides a natural mechanism to generate a approximately scale-invariant spectrum of primordial perturbations which seed structure formation, consistent with observations \cite{Planck:2018jri}. The simplest mechanism to drive inflation is a single scalar field, the inflaton, slowly rolling down its potential. The expectation of the inflationary paradigm is that sufficiently ``generic'' initial data will lead to trajectories in which the scalar field potential dominates and the universe undergoes a period of nearly exponential expansion. Since generic initial data is inhomogeneous, studying to what extent the homogeneous inflationary trajectory is an attractor requires solving the full general relativistic equations of motion.

The problem  of initial conditions for inflation has been extensively studied using analytic and semi-analytic methods~\cite{Gibbons:1977mu,Hawking:1981fz,Wald:1983ky,Starobinsky:1982mr,Barrow:1984zz,Albrecht:1984qt,Barrow:1985,Gibbons:1986xk,Jensen:1986nf,Hawking:1987bi,Penrose:1988mg,Muller:1989rp,Kitada:1991ih,Kitada:1992uh,Bruni:1994cv,Maleknejad:2012as,Gibbons:2006pa,Boucher:2011zj,Bruni:2001pc,Muller:1987hp,Barrow:1989wp,Bicak:1997ne,Capozziello:1998dq,Vachaspati:1998dy,Barrow:1987ia,Barrow:1986yf,Polyakov:2009nq,Marolf:2010nz,Tsamis:1992sx,Brandenberger:2002sk,Geshnizjani:2003cn,Marozzi:2012tp,Brandenberger:1990wu,Carroll:2010aj,Corichi:2010zp,Schiffrin:2012zf,Remmen:2013eja,Corichi:2013kua,Mukhanov:2014uwa,Remmen:2014mia,Berezhiani:2015ola,Kleban:2016sqm, Marsh:2018fsu,Finn:2018krt,Bloomfield:2019rbs, Azhar:2022yip, Albrecht:1985yf,Albrecht:1986pi,Easther:2014zga,Braden:2016tjn,Alho:2011zz,  Alho:2013vva,Brandenberger:1990xu}. In particular, early numerical relativity work on this effort was led by Goldwirth and Piran~\cite{Goldwirth:1989pr,Goldwirth:1989vz,Goldwirth:1991rj}, which focused on the spherical collapse of inhomogeneities in an inflationary theory. These simulations suggested that inflation would begin as long as there is initially a homogeneous region of radius $\sim H_0^{-1}$, where $H_0$ is the inflationary scale.

There are several different ways single-field models of inflation are classified. For example, one may classify them as high-scale or low-scale models, depending on the energy scale associated with inflation, or one may classify them as large-field or small-field models depending on their field range, and some care is needed in interpreting the different terms~\cite{Linde:2016hbb}.
For high-scale inflation, it has been argued that inflation can naturally occur under generic initial conditions \cite{Linde:1983gd,Linde:1984ir, Linde:1985ub, Linde:2014nna}, and a number of simulations have supported this claim. The first 3+1D simulations by \cite{Laguna:1991zs,KurkiSuonio:1993fg} showed that inflation could occur even if kinetic and gradient energies are initially comparable to the potential energy density. However, these simulations were limited to initial conditions that did not lead to black hole formation. More general initial conditions, including initial conditions that do lead to black hole formation, were considered in~\cite{Clough:2016ymm, Clough:2017efm, Aurrekoetxea:2019fhr, Corman:2022alv} and also showed that high-scale inflation can occur in the presence of large inhomogeneities. Some recent work has opposed this conclusion on the basis that the initial conditions and boundary choices made may bias the results towards successful inflation in some ``non-generic'' regions of low initial curvature~\cite{Garfinkle:2023vzf,Ijjas:2024oqn}. Addressing these concerns properly goes beyond the scope of this work, but we provide some brief comments in the Summary and Outlook section to set our work in context with respect to these challenges.

The robustness of low-scale models to inhomogeneous initial conditions is an even more important issue. As the energy scale is smaller, the size of the causally disconnected patches is larger, and the potential energy may be orders of magnitudes smaller than other components. Consequently, it is possible that large inhomogeneities could spoil inflation before it begins. This has motivated a wealth of studies on the effects of initial conditions on low-scale inflation, many of which involve simulating the inflating spacetime with numerical relativity in full 3+1D \cite{East:2015ggf, Clough:2016ymm, Clough:2017efm, Aurrekoetxea:2019fhr, Joana:2020rxm, Corman:2022alv, Joana:2022pzo, Joana:2024ltg}. 

In contrast to the above classifications, we follow our previous work~\cite{Aurrekoetxea:2019fhr} and classify models as models with sub-Planckian and super-Planckian characteristic scales,\footnote{The Planckian scale in the nomenclature refers to the reduced Planck scale.} depending on the scale in field space over which the potential varies appreciably. This quantity was introduced by one of the authors in~\cite{CMB-S4:2016ple}, and is a more accurate indicator both of the robustness to large inhomogeneities~\cite{Aurrekoetxea:2019fhr} and of the amount of tensor fluctuations a model produces and hence its detectability~\cite{CMB-S4:2016ple}. For a brief review see~\cite{Aurrekoetxea:2019fhr}. The simulations in~\cite{Aurrekoetxea:2019fhr} showed that for models with a sub-Planckian characteristic scale even small excursions of the field can cause inflation to fail by dragging the rest of the field to the minimum. In contrast, models with super-Planckian characteristic scales were shown to be robust to large initial gradient energies, at least for the class of initial conditions studied.  Even if the initial configuration of the field reaches the minimum, the gradient pressure of the field can pull it back into the part of the potential that supports inflation.

Here we extend the work of~\cite{Clough:2016ymm, Clough:2017efm, Aurrekoetxea:2019fhr} by investigating the effect of large and non-uniform initial inflaton velocity in addition to an inhomogeneous initial field profile. This was first studied by \cite{Corman:2022alv} for the large field case in both single and two-field inflationary models, and it was found that these models are generally robust. In our work, we consider two plateau models that are members of the $\alpha$-attractor family \cite{Kallosh:2013hoa, Kallosh:2013yoa}: the Starobinsky model~\cite{Starobinsky:1982mr} with a super-Planckian characteristic scale, and an $\alpha$-attractor model with a sub-Planckian characteristic scale. We use \textsc{grchombo}, a multipurpose numerical relativity code \cite{Andrade:2021rbd}. Our conclusions are as follows

\begin{itemize}

    \item Models with a super-Planckian characteristic scale are significantly more robust to momentum fluctuations than models with a sub-Planckian characteristic scale. For models with a sub-Planckian characteristic scale, even kinetic energy densities that are subdominant to the potential energy density can end inflation, unless the initial field configuration is confined to the plateau and sufficiently far from the minimum.
    
    \item The shorter the wavelength of the momentum fluctuation, the more energy is required for it to alter the field dynamics and end inflation. Again this is similar to what we observed for the field fluctuations \cite{Clough:2016ymm}.  So generally momentum modes with long wavelengths are more dangerous to inflation.

    \item The relative phase between the peaks of the field and momentum fluctuations significantly affects the energy required to end inflation - the closer they are aligned the less energy is needed.

\end{itemize}

The paper is organised as follows. In Section \ref{sect:theory}, we introduce the the models of inflation we consider and discuss the space of initial conditions we explore. In Section \ref{sect:results}, we discuss the numerical results. We conclude in Section \ref{sect:discussion}. We set $c=\hbar=1$ throughout, and define the (non-reduced) Planck mass $G = \mpl^{-2}$.

\section{Theory and Methodology} \label{sect:theory}

\subsection{Space of models}

We consider a model of single-field inflation with action given by
\begin{equation}
    S = \int \dd^4 x \sqrt{-g}\left(\frac{\mpl^2}{16\pi}R - \frac{1}{2}\nabla_{\mu}\phi\nabla^{\mu}\phi - V(\phi)\right)\,. \label{eqn:sf_lagrangian}
\end{equation}
We consider the $\alpha$-attractor potential \cite{Kallosh:2013hoa, Kallosh:2013yoa}
\begin{equation}
    V(\phi)=\Lambda^4\left(1-e^{\phi / \mu}\right)^2 \label{eqn:potential} \,,
\end{equation}
but we expect our results to be valid for other potentials \cite{Martin:2013tda,Martin:2024qnn}. For negative field values, the potential exponentially approaches a plateau, and the approach to the plateau is governed by the characteristic scale $\mu$. The parameter $\Lambda$ determines the energy-scale of inflation. For a given value $\mu$, $\Lambda$ is determined by the observed amplitude of the power spectrum of primordial density perturbations
\begin{equation}
\Delta_R^2=\frac{H^2}{\pi \mpl^2 \epsilon} \approx 2 \times 10^{-9} \,.
\end{equation}
Here $H$ is the Hubble parameter during inflation and $\epsilon$ is the first slow-roll parameter given by
\begin{equation}
\epsilon=\frac{\mpl^2}{16 \pi}\left(\frac{V^{\prime}}{V}\right)^2~.
\end{equation}

\subsection{Space of initial conditions} 

To investigate the effects of inhomogeneities in the scalar-field initial data on the dynamics we consider sinusoidal perturbations of the scalar field and its momentum in three spatial directions 
\begin{align}
    \phi_\mathrm{init}(\textbf{x})  & =\phi_0+\frac{\Delta \phi}{3}\sum_{i=1}^{3}\cos{\left(\frac{2\pi N_\phi x_i}{L}\right)}\,, \label{eqn:SF_profile} \\
    \Pi_\mathrm{init}(\textbf{x}) & = \Pi_{0} + \frac{\Delta \Pi}{3}\sum_{i=1}^{3}\cos{\left(\frac{2\pi N_\Pi x_i}{L} + \theta\right)} \label{eqn:SF_mom_profile}\,.
\end{align}
Here $\phi_0$ and $\Pi_{0}$ are the initial homogeneous components for the scalar field and conjugate momentum, and we impose periodic boundary conditions. $\phi_0$ was chosen such that in the absence of inhomogeneities and $\Pi_{0}=0$, inflation will last around 100 \efolds{}.\footnote{Note that this is an arbitrary choice. In models in which the ``inflationary plateau'' extends to large distances in field space, setting $\phi_0$ to be further away from the reheating region would make inflation more robust.} The values $\Delta \phi$ and $\Delta \Pi$ are the amplitudes of inhomogeneities of the scalar field and conjugate momentum. Additionally, we include a relative phase $\theta$ between the initial configurations of scalar field $\phi(x)$ and momentum $\Pi(x)$. The dimensionless wavenumbers $N_{\phi}$ and $N_{\Pi}$ determine the wavelengths of the $\phi$ and $\Pi$ fields together with the box size $L$, which we choose to be the initial Hubble length $H^{-1}_{0}$ in the absence of inhomogeneities, so that
\begin{equation}
L=\frac{3 \mpl}{\sqrt{24 \pi V\left(\phi_0\right)}} \label{eqn:hubble_radius} \,.
\end{equation}

We foliate the spacetime into spatial hypersurfaces such that the metric can be written as  
\begin{equation}
    \dd s^2 = -\alpha^2\dd t^2 + \gamma_{ij}(\dd x^i + \beta^i \dd t)(\dd x^j + \beta^j \dd t)\,.
\end{equation}
Here $\gamma_{ij}$ is the 3-dimensional spatial metric of the hypersurfaces, which we evolve along with the extrinsic curvature $K_{ij} = \partial_t \gamma_{i j}+2 D_{(i} \beta_{j)}$ as our physical degrees of freedom. The lapse function $\alpha$ and shift vector $\beta^i$ are evolved as gauge choices representing the inherent coordinate freedom of GR. Following the BSSN formulation \cite{Baumgarte:1998te, Shibata:1995we}, we decompose the extrinsic curvature into a trace $K = \gamma^{ij} K_{ij}$ and traceless part $A_{ij}$. We then extract the conformal factor $\chi$ from the spatial metric to yield the conformally related metric $\Tilde{\gamma}_{ij} = \chi \gamma_{ij}$. 

Given an initial scalar field configuration, we solve the Hamiltonian and momentum constraints of general relativity using the CTTK method \cite{Aurrekoetxea:2022mpw}. We assume an initially conformally flat metric $\tilde{\gamma}_{ij} = \delta_{ij}$ and choose $\chi=1$. Subsequently, one obtains an algebraic equation for the Hamiltonian constraint and a Poisson-like equation for the momentum constraint which can be solved to obtain the mean expansion $K$ and trace-free extrinsic curvature $A_{ij}$ respectively, see \cite{Aurrekoetxea:2022mpw} for more details. We use periodic boundary conditions, which enforces integrability conditions on the constraint equations for the initial data \cite{Aurrekoetxea:2022mpw}. That is, one requires the integral of the constraints over the domain to vanish. With an initially constant conformal factor, this leads to the condition $\int S_i \dd V = 0$, where $S_i=\Pi \partial_i\phi$ over our numerical domain. Given the sinusoidal configurations that we choose, this condition restricts the value of the phase $\theta = \{0, \pi \}$ for equal wavemodes $N_{\Pi} = N_{\phi}$.

In the FLRW limit, the trace of the extrinsic curvature $K$ is related to the Hubble parameter as $K = -3H$ and the scale factor is given by $a(t) = \chi^{-1/2}$. Given our choice of initial data, in the inhomogeneous case we define the average number of \efolds{} in terms of the conformal factor as 
\begin{equation}
    \langle\mathcal{N}\rangle=-\frac{1}{2} \langle\ln \chi\rangle \,,
\end{equation}
where the angle brackets $\langle \rangle$ denote proper volume average over the hypersurfaces defined by the $t$ coordinate 
\begin{equation}
\langle X \rangle \equiv \int_V \dd^3x \sqrt{\gamma} X ~.
\end{equation}

\subsection{Evolution and diagnostics}

We use the BSSN formulation \cite{Baumgarte:1998te,PhysRevD.52.5428,Shibata:1995we} to evolve the spacetime quantities. We employ a modified form of moving puncture gauge  \cite{Campanelli:2005dd,Baker:2005vv}. We evolve the lapse according to
\begin{equation}
\partial_t \alpha = - \alpha e^{- \alpha} K+\beta^i \partial_i \alpha \,,
\end{equation}
and invoke the standard Gamma-driver condition for the shift. 

The exponential damping prevents the lapse from growing too large for expanding regions (i.e. where $K < 0$), but approaches the standard 1+log form for collapsing regions.

\begin{figure}[t!]
\centering
\includegraphics[width=1.0\columnwidth]{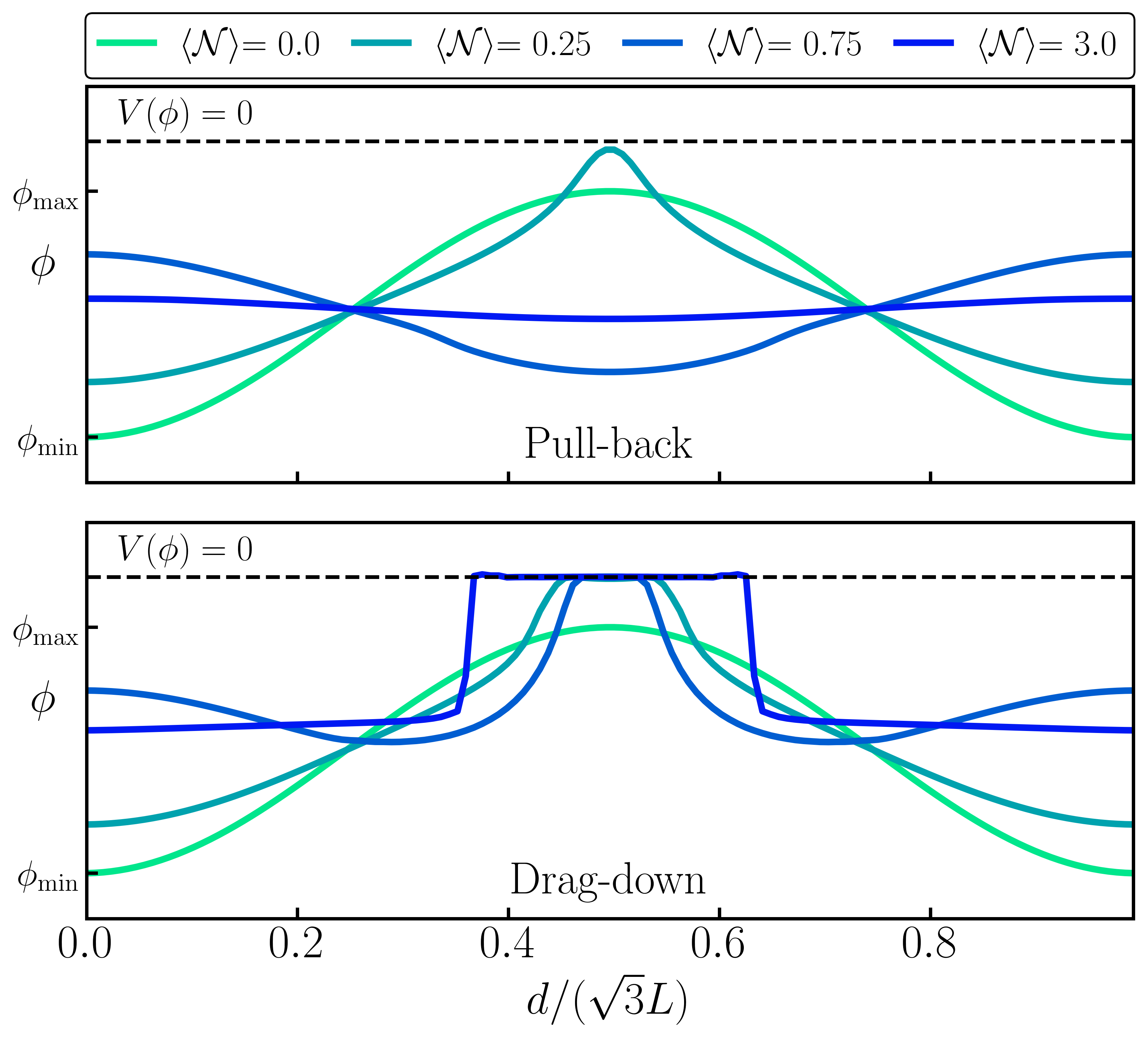}
\caption{Evolution of the scalar field across the largest diagonal of our numerical domain $(0,0,0)\rightarrow (L,L,L)$. Both panels start from the same scalar field configuration $N_\phi=N_{\Pi}=1$ and phase $\theta = 0$, but different $\Delta\Pi$. The top panel shows the pull-back, when the scalar field homogenizes. The bottom panel shows the effect of  larger initial momentum, which can push the field sufficiently down the potential to remain stuck at the minimum (dashed) and drag down the rest with it.}
\label{fig:pull_drag}
\end{figure}

\begin{figure}[t!]
\centering
\includegraphics[width=1.0\columnwidth]{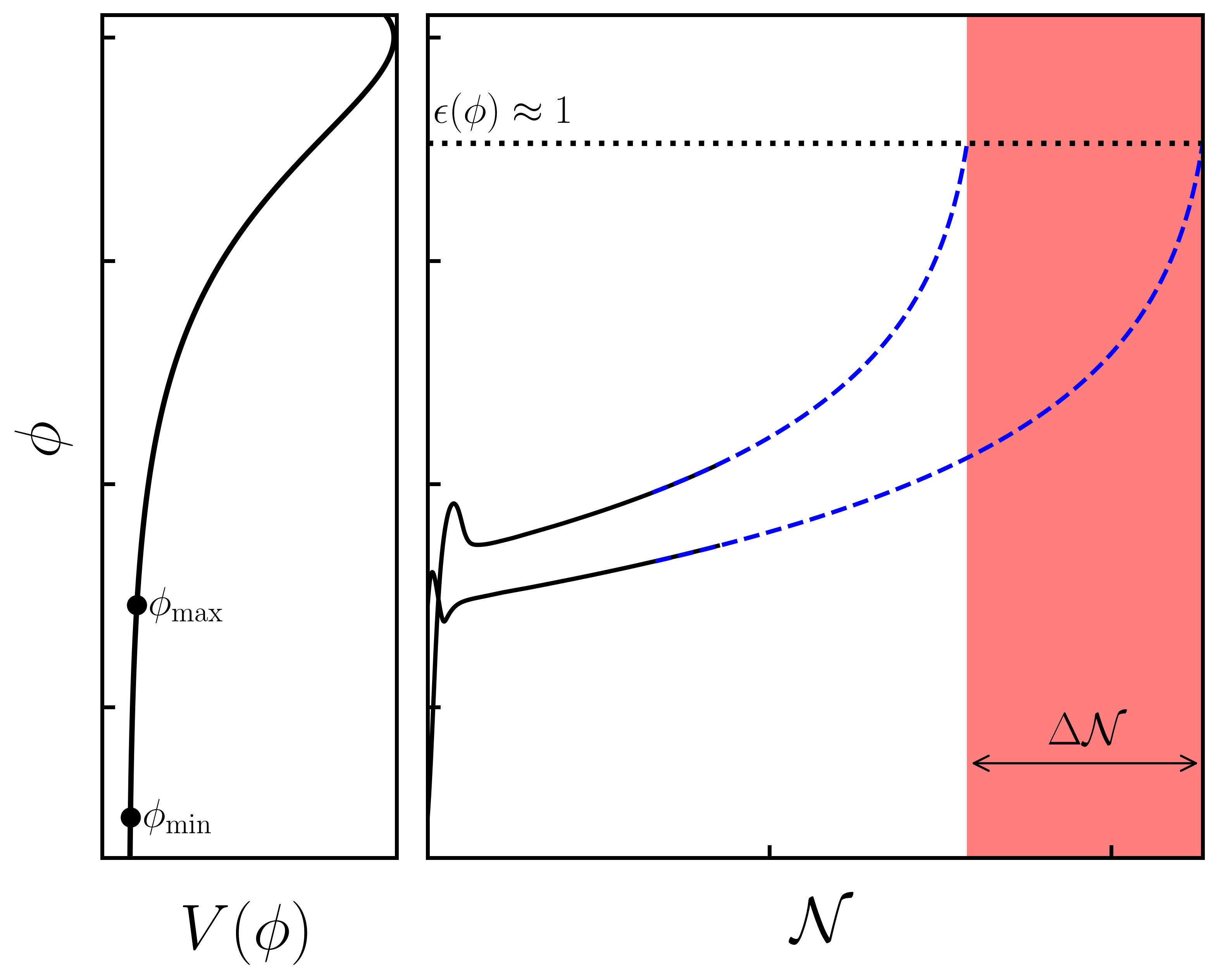}
\caption{Left: illustration of the $\alpha$-attractor potential and and initial minimum $\phi_{\mathrm{min}}$ and maximum $\phi_{\mathrm{max}}$ values of the scalar field. Right: evolution of $\phi_{\mathrm{min}}$ and $\phi_{\mathrm{max}}$ with the local number of \efolds{} $\mathcal{N}$ for the entirety of inflation. The values in black lines are extracted from numerical relativity simulations. As the field homogenizes significantly after a few \efolds{}, we evolve the field extrema to the end of slow-roll $\epsilon \approx 1$ via Eqn. (\ref{eq:hom_KG}-\ref{eq:hom_FRW}) (dashed blue lines). We use this to predict the range of expected number of \efolds{} $\Delta \mathcal{N}\equiv\mathcal{N}_\mathrm{min}-\mathcal{N}_\mathrm{max}$. }
\label{fig:potential}
\end{figure}

In \cite{Aurrekoetxea:2019fhr} we described two outcomes when evolving a scalar field configuration: (i) \textit{pull-back} when the field restores itself to a more homogeneous configuration and (ii) \textit{drag-down} when some part of the field rolls down the potential and drags the rest of the field configuration with it, preventing inflation from starting. In the absence of initial momentum, we found that for inflationary models with sub-Planckian characteristic scales there exists a critical amplitude beyond which the field is dragged down $\Delta\phi>\Delta\phi_\mathrm{crit}$. In this paper we fix the inhomogeneous scalar field configuration $\phi(x)$ and study the impact of varying its initial conjugate momentum $\Pi(x)$. In Fig. \ref{fig:pull_drag} we show two example simulations, one in which the field is pulled back (top panel) and one in which the field is dragged down to the minimum of the potential as $\Pi(x)$ is increased (bottom panel). Most of the dynamics occurs within the first few \efolds{}, after which the inhomogeneities exit the horizon (if inflation gets started), and subsequently slowly rolls down the potential (black line in Fig. \ref{fig:potential}). 

A typical simulation leads to disjoint regions of space, with some regions where the scalar field has reached the reheating minimum (ending inflation) and some regions in which the scalar field is approximately homogeneous and on the inflating plateau. In the regions in which the field reaches the minimum, the field oscillates. Because of the much shorter timescale of this oscillation, we rapidly lose accuracy and hence we assumed that such regions are no longer inflating. For regions in which the field remains in the plateau, inflation persists, although in each such region the field may take different (approximately homogeneous) field values.

To estimate the expected number of \efolds{} in these latter regions, we extend the evolution of $\phi_\mathrm{max}$ and $\phi_\mathrm{min}$ by solving the \emph{homogeneous} 
 evolution equations
\begin{align}
    \ddot{\phi} &+3H\dot{\phi} + V'(\phi) = 0\,, \label{eq:hom_KG} \\
    \left( \frac{\dot{a}}{a} \right)^2 & = \frac{8\pi }{3\mpl^2}\left(\frac{1}{2}\dot{\phi}^2 + V(\phi)\right)\,,\label{eq:hom_FRW}
\end{align}
where we take as initial conditions the values of $\phi$ and $\dot{\phi}$ extracted from numerical relativity simulations.  We can use the homogeneous equations as these regions have spatial dimensions equal to or exceed the Hubble radius associated with the inflationary vacuum  $\geq H_{0}^{-1}$. We have checked in \cite{Aurrekoetxea:2019fhr} that this gives excellent agreement with numerical results.

The result is shown in Fig. \ref{fig:potential}, where we plot the evolution of the maximum and minimum values of the scalar field (dashed blue lines), extending the results from numerical relativity simulations (solid black lines). This allows us to define a range of expected number of \efolds{} as $\Delta\mathcal{N}\equiv \mathcal{N}_\mathrm{max} - \mathcal{N}_\mathrm{min}$.

Writing the Hamiltonian constraint as
\begin{equation}
    K^2 = \frac{3}{2}\Big(16\pi \rho + A_{ij}A^{ij} - R\Big) \label{eqn:K-Ham} \,.
\end{equation}
and noting that for a spatially homogeneous and isotropic universe, 
 $A_{ij}A^{ij}$ vanishes, the Friedmann equation is recovered via the definitions $K = -3H$ and $R = 6k/a^2$ where $k=0$ in our case. Splitting the scalar-field energy density $\rho$ into its respective components and taking the volume average yields 
\begin{equation}
    1 = \Omega_\mathrm{pot} + \Omega_\mathrm{grad} + \Omega_\mathrm{kin} + \Omega_{A_{ij}} + \Omega_R \label{eqn:Omega-Ham} \,,
\end{equation}
where we have defined the density parameters
\begin{equation}
    \Omega_{i} = c_i\frac{\langle \rho_{i} \rangle}{\langle K^2 \rangle} \,,
\end{equation}
for the respective densities $\rho_i$, including the potential $ V\left(\phi\right)$ and curvature terms $A_{ij}A^{ij}$ and $R$. Here $c_i$ are the associated factors i.e. $c_i = 24 \pi$ for the matter density terms and $c_i=3/2$ for the curvature terms.

As long as the potential energy dominates ($\Omega_\mathrm{pot}\approx 1$) over a large enough volume and sufficient time, the exponential expansion irons out any inhomogeneities and the region approaches the quasi-de Sitter solution.  We vary the initial gradient and kinetic energies by choosing different profiles of the field $\phi(x)$ and conjugate momentum $\Pi(x)$. Solving the Hamiltonian and momentum constraints sources additional shear contributions via $A_{ij}A^{ij}$, which \emph{a priori} are not known. For this reason, we parameterize the amplitude of inhomogeneities in terms of the departure from potential energy domination $(1-\Omega^0_\mathrm{pot})$, the total amount of energy densities working against inflation. We also track the evolution of diagnostic quantities constructed from the Weyl tensor, the Weyl curvature and Chern-Pontryagin invariants, see \cite{Ijjas:2023bhh} for more details.

\begin{figure*}[t!]
\centering
\includegraphics[width=1.0\textwidth]{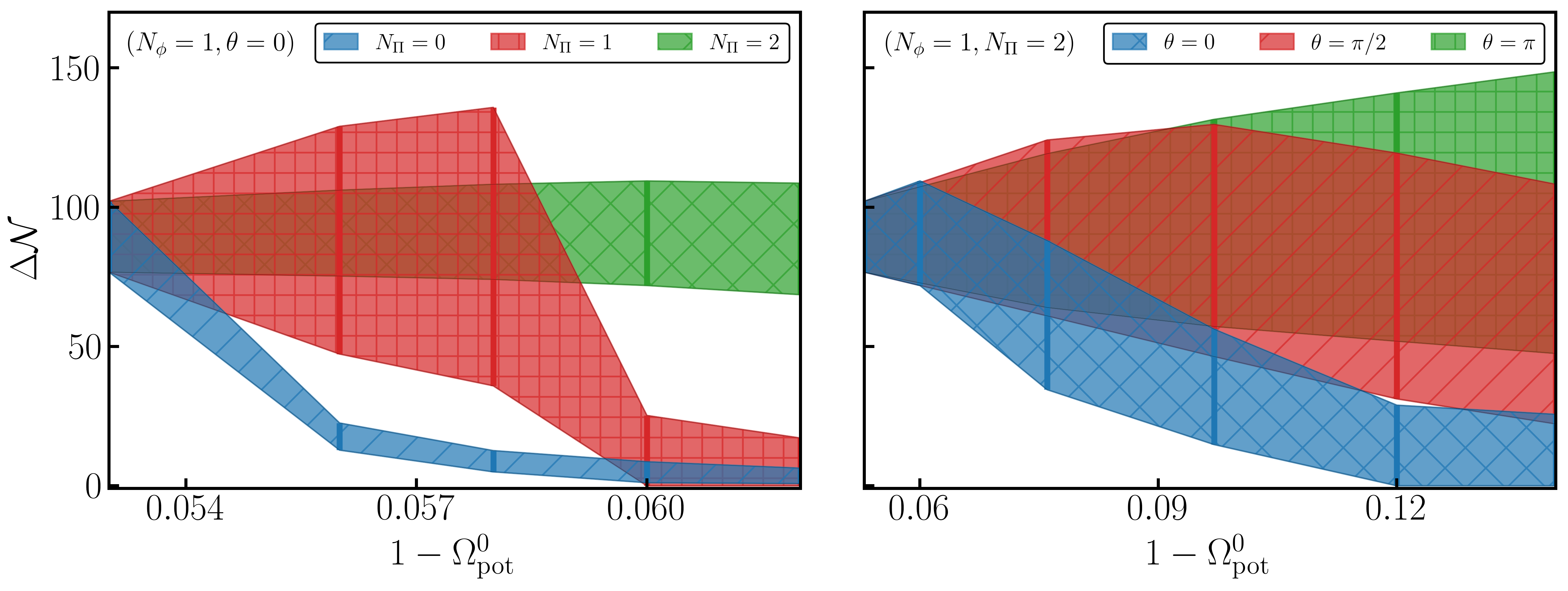}
\caption{The effect of varying the field momentum for different wavelengths ($N_\Pi$) and relative phases ($\theta$) with the scalar field configuration on the expected number of \efolds{} $\Delta\mathcal{N}$ for the case with sub-Planckian characteristic scale. Left: We fix the relative phase $\theta = 0$ and study the effect of different wavelengths $N_\Pi$. The lowest mode $N_{\Pi} = 0$ (uniform kick) is the most dangerous case, quickly reducing $\Delta\mathcal{N}$. The higher the wavelength, the more robust the configuration. Right: We fix the wavelength  $N_{\Pi} = 2$ and study the effect of relative phases $\theta$ between the field and momentum. The configuration for which the maxima are aligned ($\theta = 0$ for $(N_\phi,N_\Pi)=(1,2)$) is the most dangerous case.}
\label{fig:A_r_efolds}
\end{figure*}

\section{Numerical results} \label{sect:results}

\subsection{Sub-Planckian characteristic scale}

For the model with sub-Planckian characteristic scale, we fix the model parameters so that $\mu = 5\times 10^{-4}\mpl$ and $\Lambda^4 = 1.18\times 10^{-18}\mpl^4$. We also fix the initial scalar field configuration choosing $\phi_0 = -6.33\times 10^{-2}\mpl$, $\Delta\phi = 4.5\times 10^{-2}\mpl$ and $N_\phi=1$. We vary parameters of the initial conjugate momentum: the amplitude $\Delta \Pi$, mode $N_{\Pi}$ and relative phase $\theta$, which generate configurations with different initial $(1-\Omega^0_\mathrm{pot})$. The precise values are \emph{a priori} not known as they depend on the solution of the Hamiltonian and momentum constraints for given $\phi^0(x)$ and $\Pi^0(x)$ configurations. However, for models with sub-Planckian characteristic scale, even small gradient and kinetic energy densities prevent inflation from starting. They do not source a significant shear contribution via $A_{ij}A^{ij}$, so that $1-\Omega^0_\mathrm{pot}\approx \Omega^0_\mathrm{grad}+\Omega^0_\mathrm{kin}\ll 1$. 

In agreement with our previous studies that did not include initial momentum \cite{Clough:2016ymm, Aurrekoetxea:2019fhr}, we find that models with sub-Planckian characteristic scale are not robust and fail to provide a sufficient number of \efolds{} even when the potential dominates initially $\Omega^0_\mathrm{pot}\approx 1$. Adding momentum does not change the overall conclusion that sub-Planckian models are susceptible to inhomogeneities. In Fig. \ref{fig:A_r_efolds} we plot how the expected number of \efolds{} $\Delta\mathcal{N}$ varies as we decrease ($1-\Omega^0_\mathrm{pot}$). 

\emph{In-phase perturbations}. In the left panel we fix the relative phase $\theta=0$ and study the impact of increasing $(\Pi_0,\Delta\Pi)$ for three dimensionless wavenumbers $N_\Pi=\{0,1,2\}$ of the scalar momentum. The $N_\Pi=0$ case exhibits the most effective reduction in \efolds{}, as the field is uniformly boosted towards the potential minimum. For $N_\Pi=1$ we observe an initial increase in the maximum number of \efolds{} as the kinetic energy is increased. This is because the momentum fluctuation is in phase with the field fluctuation and at late times one finds an even larger spread of the field. However, increasing the kinetic energy eventually prevents the pull-back of the field up the plateau by gradients. Subsequently, the field rolls down to the potential minimum ending inflation prematurely. Lastly, we find that for $N_{\Pi} = 2$  there is no significant change in $\Delta\mathcal{N}$ for the range of energies plotted in the left panel.  

\emph{Out-of-phase perturbations}. In the right panel we focus on the most robust wavelength we have simulated ($N_\Pi=2$) and study the impact of varying the relative phase $\theta$ between the field $\phi$ and momentum $\Pi$ configurations. This results in different momentum density profiles and amplitudes $S_i\sim \Pi \partial_i\phi$. The relative phase  of the initial $\phi$ and $\Pi$ profiles significantly impact the likeliness of failure.  We find that if the perturbations of $\Pi$ and $\phi$ are out of phase, then the effects can even cancel each other, resulting in more robust inflation with an number of \efolds{} that can be larger than for the homogeneous initial conditions.

We conclude that modes with larger wavenumbers result in a larger expected number of \efolds{} and thus, similarly to perturbations of the scalar amplitude \cite{Clough:2016ymm,Aurrekoetxea:2019fhr}, \textit{longer wavelength kinetic inhomogeneities are the most dangerous}. This agrees with what one might expect as smaller wavelength momentum modes source shorter wavelength inhomogeneities in the scalar field, which experience larger gradient forces that subsequently pull back and homogenize the field. While the larger gradients source larger curvatures, at least locally, the primary dynamic that determine inflationary failure is the motion of the scalar field (as opposed to dynamics of the curvature), which is consistent with our previous results. In light of this, our results unsurprisingly show that the most dangerous phases are those for which the field and momentum maxima are in phase.

\subsection{Super-Planckian characteristic scale}

Previous studies \cite{Clough:2016ymm,Aurrekoetxea:2019fhr,East:2015ggf} showed that large field inflation is robust to large initial scalar field inhomogeneities. Furthermore, in \cite{Corman:2022alv}, it was also shown that this resilience carries through to large initial scalar field momenta\footnote{See also \cite{Albrecht:1986pi} for an early investigation.}. In this section, we confirm the latter results.

As a concrete example, we study the robustness of Starobinsky inflation, for which $\mu = \sqrt{3/16\pi} \mpl $ and $\Lambda^4 = 2.64 \times 10^{-15}\mpl^4$. Given the results from simulations with sub-Planckian characteristic scale, we only focus on the most dangerous scenario: uniform kicks ($N_\Pi=0$) towards the potential minimum. We compare this effect for two scalar field configurations with $\phi_0/\mpl =-1.2$ and $\Delta\phi/\mpl= \lbrace 0.19, 0.38 \rbrace$, resulting in gradient energies densities $\rho^0_\mathrm{grad}\approx V(\phi)$ and $\rho^0_\mathrm{grad}\approx 3V(\phi)$ respectively. 
Note that the  initial density parameter $1-\Omega^0_\mathrm{pot}\approx \Omega^0_\mathrm{grad}+\Omega^0_\mathrm{kin}+\Omega^0_{A_{ij}}$ is a priori unknown before solving the constraints, because for models with a super-Planckian characteristic scale the shear contribution given by $A_{ij}A^{ij}$ is significant. This shear contribution limits the range of initial conditions that we can explore, as it results in collapsing regions in which the evolution breaks down. We believe that the breakdown at these points is not physical and is instead related to numerical problems in evolving collapsing planar gravitational wave structures. Such problems have been seen elsewhere in critical collapse simulations of gravitational waves, and while progress has been made, such scenarios remain challenging to evolve in the standard gauges that we currently use \cite{Hilditch:2013cba}. We hope to employ the recent advances in critical collapse simulations \cite{Baumgarte:2022ecu,Cors:2023ncc,Baumgarte:2023tdh} in future work, as discussed further in the Outlook section below.

\begin{figure}[t!]
\centering
\includegraphics[width=1.0\columnwidth]{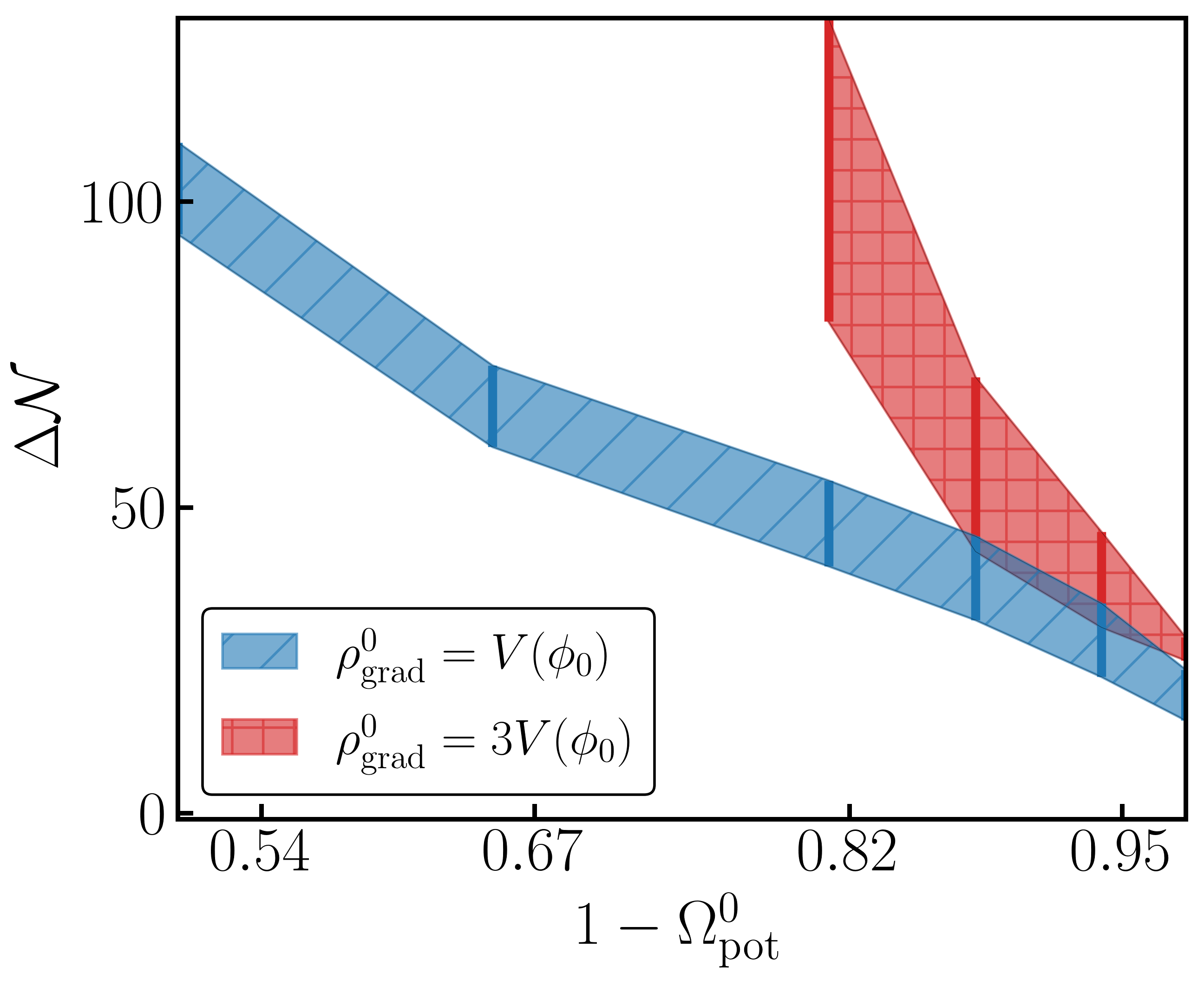}
\caption{The effect of increasing the momentum (decreasing $\Omega^0_\mathrm{pot}$) on the range of the expected number of \efolds{} $\Delta\mathcal{N}$ for the Starobinsky model (super-Planckian characteristic scale). We show two sets of runs with different initial gradient energy densities $\rho_\mathrm{grad}$  beginning with $\Pi=0$. We conclude that Starobinsky inflation is robust, as they can support significant inhomogeneities even when the initial potential energy is subdominant $\Omega^0_\mathrm{pot}< 0.5$.}
\label{fig:LF_efolds}
\end{figure}

In Fig. \ref{fig:LF_efolds}, we plot the reduction in the expected number of \efolds{} $\Delta\mathcal{N}$ as the momentum is increased (and $\Omega^0_\mathrm{pot}$ is reduced). In contrast to the model with sub-Planckian characteristic scale, we observe significant robustness even to strong uniform boosts for which the potential vacuum energy is highly subdominant $\Omega^0_\mathrm{pot}\approx 0.05$. So models of inflation with a super-Planckian characteristic scale are robust even to large initial momenta, consistent with the results of \cite{Corman:2022alv}. Moreover, we find that for fixed $1-\Omega^0_\mathrm{pot}$, \emph{somewhat counter-intuitively larger gradient energy densities make the model more robust} -- again this was first noted by \cite{Corman:2022alv}. We agree with their argument that this is a consequence of higher gradient restoring forces, which are more efficient in homogenizing the scalar field.

We conclude that large fluctuations in the conjugate momentum are more dangerous than large perturbations in the scalar field itself. This may not be surprising as the restoring forces from scalar gradients act to homogenize the configuration to the mean value, whereas the conjugate momentum actively pushes the field to a more inhomogeneous distribution, and possibly towards the potential minimum to end inflation.

\section{Summary and Outlook}\label{sect:discussion}

\begin{figure}[t!]
\centering
\includegraphics[width=\columnwidth]{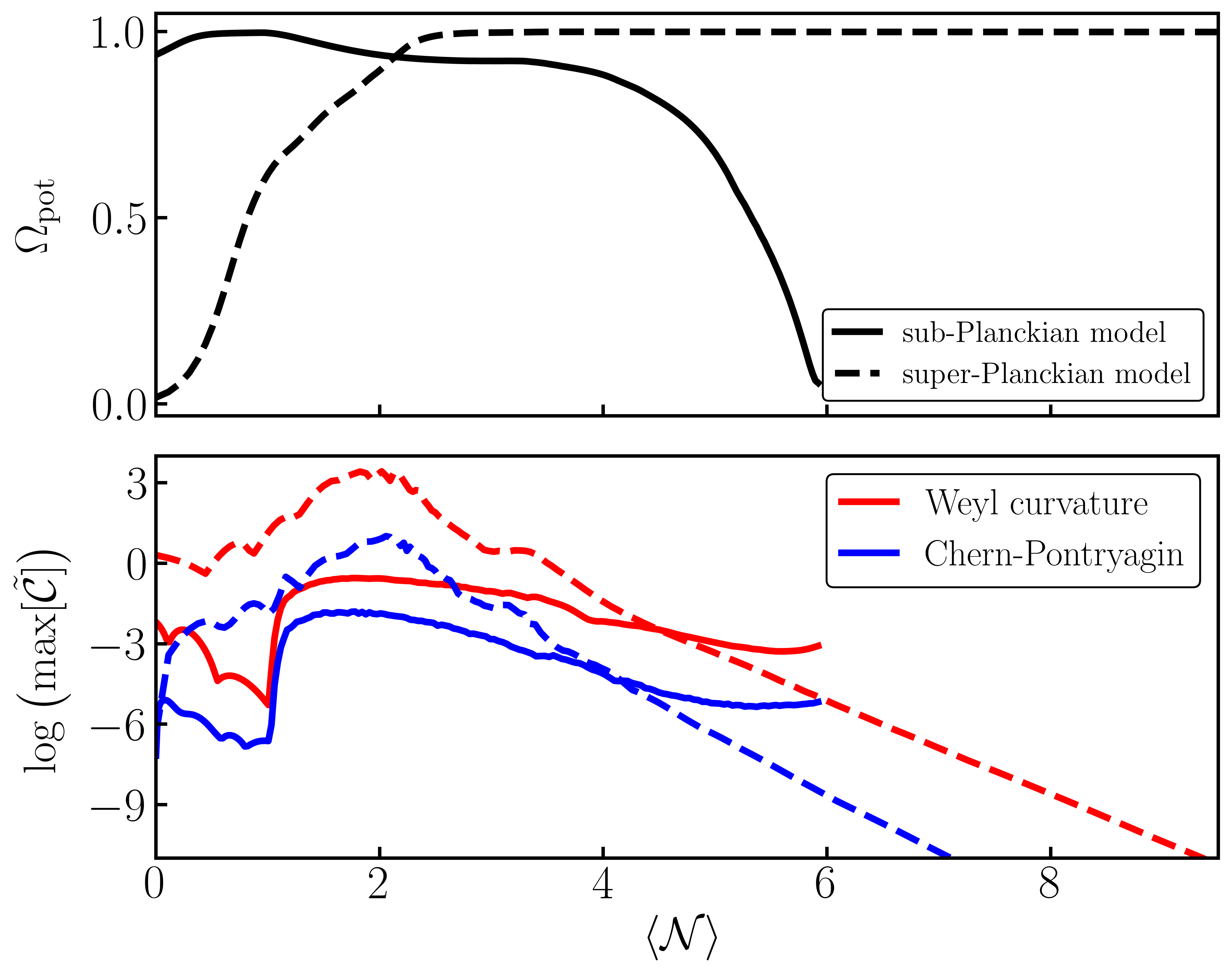}
\caption{Top: Evolution of the potential density parameters for models with sub-Planckian (solid) and super-Planckian (dashed) characteristic scales. Bottom: The evolution of the maximum values of the Weyl (red) and Chern-Pontryagin (blue) curvature scalars. Both panels show the failure of inflation for the model with sub-Planckian characteristic scale even when the potential initially dominates and the initial curvature scalars are small. On the other hand, the model with super-Planckian characteristic scale (Starobinsky inflation) is successful in starting inflation and homogenizing the field for subdominant potential energy densities and in the presence of $\mathcal{O}(1)$ curvature invariants.}
\label{fig:densities_curvatures_efolds}
\end{figure}

We have studied the robustness of inflation to large inhomogeneities in the scalar field and its conjugate momentum. In general, introducing momentum fluctuations in the scalar field makes inflation more likely to fail, especially for low frequency perturbations. Furthermore, the relative phase between the scalar field and conjugate momentum plays an important role. The case in which the phases of the inhomogeneities in field and in the momentum are aligned is the most likely to result in failure.  Nevertheless, the introduction of momentum fluctuations does not significantly alter the evolution from that of a purely inhomogeneous field configuration. 
The results are consistent with our previous work which suggests it is scalar field dynamics, i.e. how the field is being pushed to the reheating zone via the initial inhomogeneities, that determines the robustness of inflation. On the other hand, we find that a combination of both kinetic and scalar field perturbations can interact in interesting ways. For example, we find that for fixed $\Omega^0_\mathrm{pot}$, increasing the gradient energy density makes inflation more robust. That is, departures from $\Omega^0_\mathrm{pot} = 1$ are more robust if they are contained in $\Omega^0_\mathrm{grad}$ than in $\Omega^0_\mathrm{kin}$. 

Previous results \cite{East:2015ggf, Clough:2016ymm,Aurrekoetxea:2019fhr, Joana:2020rxm,Corman:2022alv} have shown that models with super-Planckian characteristic scale are robust to initial perturbations, and we find that this broad conclusion remains true in the presence of large kinetic fluctuations. Models of inflation with sub-Planckian characteristic scale remain susceptible to a variety of initial inhomogeneities. Even though we have focused on two particular cases of $\alpha$-attractor inflation, we expect these results to apply more generally.  In Fig. \ref{fig:densities_curvatures_efolds} we illustrate this, plotting the evolution of the potential density parameter in the top panel and the maximum values of the (rescaled) Weyl curvature scalar
\begin{equation}
    \mathcal{\tilde{C}}_{\mathrm{W}} = \frac{\mathcal{C}^{\mu \nu \rho \omega}\mathcal{C}_{\mu \nu \rho \omega}}{\langle K^{4} \rangle}  \,, 
\end{equation}
and Chern-Pontryagin invariant 
\begin{equation}
\mathcal{\tilde{C}}_{\mathrm{P}} = \frac{{}^{*}\mathcal{C}^{\mu \nu \rho \omega}\mathcal{C}_{\mu \nu \rho \omega}}{\langle K^{4} \rangle
} \,,
\end{equation}
in the bottom panel. Here $\mathcal{C}_{\mu \nu \rho \omega}$ is the Weyl curvature tensor and ${}^{*}\mathcal{C}^{\mu \nu \rho \omega}$ its dual. Employing such curvature invariants was suggested as a diagnostic to determine the success of inflation in  \cite{Ijjas:2023bhh}.

Models with sub- and super-Planckian characteristic scales are denoted by the solid and dashed lines, respectively. We find that while models with sub-Planckian characteristic scale can fail even when the potential density parameter dominates $\Omega^0_\mathrm{pot}\approx 1$ and the curvature invariants are small, models with super-Planckian characteristic scale show greater robustness, homogenizing the spacetime even starting from large departures with $\Omega^0_\mathrm{pot}\approx \mathcal{O}(0.01)$. 
We see that for a super-Planckian characteristic scale, we can drive the Weyl curvature invariants to subdominant values. The case we study appears roughly comparable to the ``best'' inflation case studied in \cite{Ijjas:2024oqn}, with maximum curvature scalars that take O(1) values during the simulation. As discussed in \cite{Garfinkle:2023vzf}, this finding does not address the question of whether the inflation that we observe is being dominated by ``non generic'' initial regions with lower values of the curvature - to do this would require a more accurate tracking of observers through the evolution to map initial to final regions. As in \cite{Ijjas:2024oqn}, we are limited in our choice of initial data to zero average spatial curvature, and as a result even the maximum initial values of the Chern-Pontryagin invariant are small.

The strong dependence of our results on the characteristic scale are particularly noteworthy because upcoming CMB experiments are designed to detect or exclude the imprint of primordial gravitational waves on the CMB for models of inflation with a super-Planckian characteristic scale~\cite{CMB-S4:2016ple}. As a consequence the absence of a detection may disfavor the models that are more robust to inhomogeneities. In the context of inflation, this might suggest the need for some mechanism that sets up initial conditions that are suitable for the models with sub-Planckian characteristic scale. Such a mechanism could be as simple as an earlier period of inflation. Here we have studied a model with a super-Planckian characteristic scale, and one with a characteristic scale that is well below the Planck scale. We will sharpen these statements by studying the transition in more detail in future work.

Our work so far has focused on conformally flat initial conditions and it would be natural to relax this assumption in future work. In addition, one could try to avoid the use of periodic boundary conditions altogether, and the resulting constraints on initial data, by using spherically adapted coordinates that asymptote to a specific cosmology. These changes would potentially permit the study of larger initial curvature invariants, which would provide more stringent tests of models with a  super-Planckian characteristic scale. More work could also consider how superhorizon modes, as part of a power spectrum of initial perturbations, can affect the overall robustness.

The effect of gauge also merits further study. In our simulations, the foliation of space-time is decided by a choice of the lapse $\alpha$ and shift $\beta_i$ parameters, which are dynamically evolved using an adapted form of the moving puncture gauge \cite{Campanelli:2005dd,Baker:2005vv}.
We have detailed above the challenges this introduces when we follow the evolution through multiple \efolds{} in regions with very different expansion rates, and the instabilities that appear to arise with the collapse of strong tensor perturbations in the larger field case. In \cite{Garfinkle:2023vzf, Ijjas:2023bhh,Ijjas:2018cdm}, a different choice of gauge corresponding to constant mean curvature slices is employed, which enable the authors to keep track of the evolution with large numbers of \efolds{}. Other ideas for improved gauge conditions have also been proposed in \cite{Doniere:2023ebv,Baumgarte:2022ecu}. It would be valuable to explore whether such strategies can also be exploited or adapted to the BSSN formalism. 

\acknowledgments
We acknowledge useful conversations with Maxence Corman, William East, Alan Guth, David Kaiser and Sonia Paban. We would also like to thank the GRChombo team \href{http://www.grchombo.org}{(http://www.grchombo.org/)}. This work is supported by a Research Project Grant RPG-2021-423 from Leverhulme Trust. JCA acknowledges funding from the Beecroft Trust and The Queen’s College via an extraordinary Junior Research Fellowship (eJRF). ME has been supported in part by the PID2021-123703NB-C21 grant funded by MCIN/AEI/10.13039/501100011033/and by ERDF; “A way of making Europe”; the Basque Government grant (IT-1628-22). RF was supported in part by the Department of Energy under Grant No. DE-SC0009919.

This work was performed using the DiRAC@Durham
facility managed by the Institute for Computational
Cosmology on behalf of the STFC DiRAC HPC Facility (www.dirac.ac.uk) under DiRAC RAC13 Grant
ACTP238 and DiRAC RAC15 Grant ACTP316. The
equipment was funded by BEIS capital funding via
STFC capital grants ST/P002293/1, ST/R002371/1 and
ST/S002502/1, Durham University and STFC operations
grant ST/R000832/1. This work also used the DiRAC
Data Intensive service at Leicester, operated by the University of Leicester IT Services, which forms part of the STFC DiRAC HPC Facility (www.dirac.ac.uk). The equipment was funded by BEIS capital funding via STFC capital grants ST/K000373/1 and ST/R002363/1 and STFC DiRAC Operations grant ST/R001014/1. DiRAC
is part of the National e-Infrastructure.

\bibliography{mybib.bib}
\clearpage

\appendix

\section*{Summary of simulations and code validation}\label{appendix}

We set the length of our computational domain $L = 32 M$ and impose periodic boundary conditions. Our standard runs have $N^3=128^3$ number of coarse grid points of the model with sub-Planckian characteristic scale and $N^3=256^3$ for the super-Planckian case. In Table \ref{tab:model_params} we list the summary of simulations. We fix $\phi_0 = -6.33\times 10^{-2}\mpl$ and $\Delta\phi = 4.5\times 10^{-2}\mpl$ for sub-Planckian runs, and $\phi_0 =-1.2\mpl$ and $\Delta\phi= \lbrace 0.19\mpl, 0.38\mpl \rbrace$ for super-Planckian.

\begin{table}[b!]
Sub-Planckian runs:
\begin{tabular}{|c|c|c|c|c|c|}
\hline
Run & $N_{\Pi}$ & $\theta$ & $\Pi_0$  [$\mpl^2$] & $\Delta \Pi $ [$\mpl^2$]   &  $1 - \Omega^0_{\mathrm{pot}}$  \\
\hline
\hline
SF$01$ & $0$ & N/A & $0$ & $0$ & $5.30 \times 10^{-2}$ \\
\hline
SF$02$ & $0$ & N/A & $7.68 \times 10^{-11}$ & $0$ & $5.52 \times 10^{-2}$  \\
\hline
SF$03$ & $0$ & N/A & $1.09 \times 10^{-10}$ & $0$ & $5.75 \times 10^{-2}$ \\
\hline
SF$04$ & $0$ & N/A & $1.33 \times 10^{-10}$ &$0$ & $5.97 \times 10^{-2}$  \\
\hline
SF$05$ & $0$ & N/A  & $1.54 \times 10^{-10}$ & $0$ & $6.20 \times 10^{-2}$   \\
\hline
\hline
SF$11$ & $1$ & $0$ & $0$ & $1.88 \times 10^{-10}$ & $5.52 \times 10^{-2}$ \\
\hline
SF$12$ & $1$ & $0$ & $0$ & $2.77 \times 10^{-10}$ & $5.74 \times 10^{-2}$ \\
\hline
SF$13$ & $1$ & $0$ & $0$  & $3.26 \times 10^{-10}$ & $5.97 \times 10^{-2}$\\
\hline
SF$14$ & $1$ & $0$ & $0$ & $3.76 \times 10^{-10}$ & $6.19 \times 10^{-2}$ \\
\hline
\hline
SF$21$ & $2$ & $0$ & $0$ & $5.95 \times 10^{-10}$ & $7.51\times 10^{-2}$\\
\hline
SF$22$ & $2$ & $0$ & $0$  & $8.41 \times 10^{-10}$ & $9.66 \times 10^{-2}$ \\
\hline
SF$23$ & $2$ & $0$ & $0$ &$1.03 \times 10^{-9}$ & $1.17 \times 10^{-1}$ \\
\hline
SF$24$ & $2$ & $0$ & $0$  & $1.19 \times 10^{-9}$ & $1.37 \times 10^{-1}$\\
\hline
\hline
SF$21\pi$ & $2$ & $\pi$ & $0$ &$5.95 \times 10^{-10}$ & $7.51\times 10^{-2}$ \\
\hline
SF$22\pi$ & $2$ & $\pi$ & $0$ & $8.41 \times 10^{-10}$ & $9.66 \times 10^{-2}$\\
\hline
SF$23\pi$ & $2$ & $\pi$ & $0$  & $1.03 \times 10^{-9}$ & $1.17 \times 10^{-1}$ \\
\hline
SF$24\pi$ & $2$ & $\pi$ & $0$  & $1.19 \times 10^{-9}$ & $1.37 \times 10^{-1}$ \\
\hline
\hline
SF$21\pi2$ & $2$ & $\pi/2$ & $0$ & $5.95 \times 10^{-10}$ & $7.51\times 10^{-2}$  \\
\hline
SF$22\pi2$ & $2$ & $\pi/2$ & $0$ & $8.41 \times 10^{-10}$ & $9.66 \times 10^{-2}$\\
\hline
SF$23\pi2$ & $2$ & $\pi/2$ & $0$  & $1.03 \times 10^{-9}$ & $1.17 \times 10^{-1}$ \\
\hline
SF$24\pi2$ & $2$ & $\pi/2$ & $0$ & $1.19 \times 10^{-9}$ & $1.37 \times 10^{-1}$ \\
\hline
\end{tabular}

\vspace{12pt}
Super-Planckian runs:\\
\begin{tabular}{|c|c|c|c|}
\hline
Run & $\Delta \phi$ [$\mpl$] & $\Pi_0$  [$\mpl^2$] &  $1 - \Omega^0_{\mathrm{pot}}$ \\
\hline
\hline
LF$11$ & $1.90 \times 10^{-1}$ & 0 & $5.04 \times 10^{-1}$ \\
\hline
LF$12$ &   $1.90 \times 10^{-1}$&   $1.67 \times 10^{-7}$ & $6.45 \times 10^{-1}$ \\
\hline
LF$13$ &   $1.90 \times 10^{-1}$ & $3.34 \times 10^{-7}$ & $8.10 \times 10^{-1}$\\
\hline
LF$14$ &   $1.90 \times 10^{-1}$&  $4.68 \times 10^{-7}$  &$8.81 \times 10^{-1}$ \\
\hline
LF$15$ &   $1.90 \times 10^{-1}$ &  $7.35 \times 10^{-7}$  & $9.44 \times 10^{-1}$ \\
\hline
LF$16$ &   $1.90 \times 10^{-1}$ & $1.27 \times 10^{-6}$ & $9.80 \times 10^{-1}$ \\
\hline
\hline
LF$21$  & $3.80 \times 10^{-1}$ & $0$ &$8.05 \times 10^{-1}$ \\
\hline
LF$22$  & $3.80 \times 10^{-1}$  & $2.67 \times 10^{-7}$ & $8.78 \times 10^{-1}$\\
\hline
LF$23$  & $3.80 \times 10^{-1}$ & $5.34 \times 10^{-7}$ & $9.45 \times 10^{-1}$ \\
\hline
LF$24$  &$3.80 \times 10^{-1}$& $1.07 \times 10^{-6}$ & $9.83 \times 10^{-1}$ \\
\hline
\end{tabular}
\caption{Summary of our simulations with sub-Planckian and super-Planckian characteristic scales.}
\label{tab:model_params}
\end{table}

\begin{figure}[t!]
\centering
\includegraphics[width=\columnwidth]{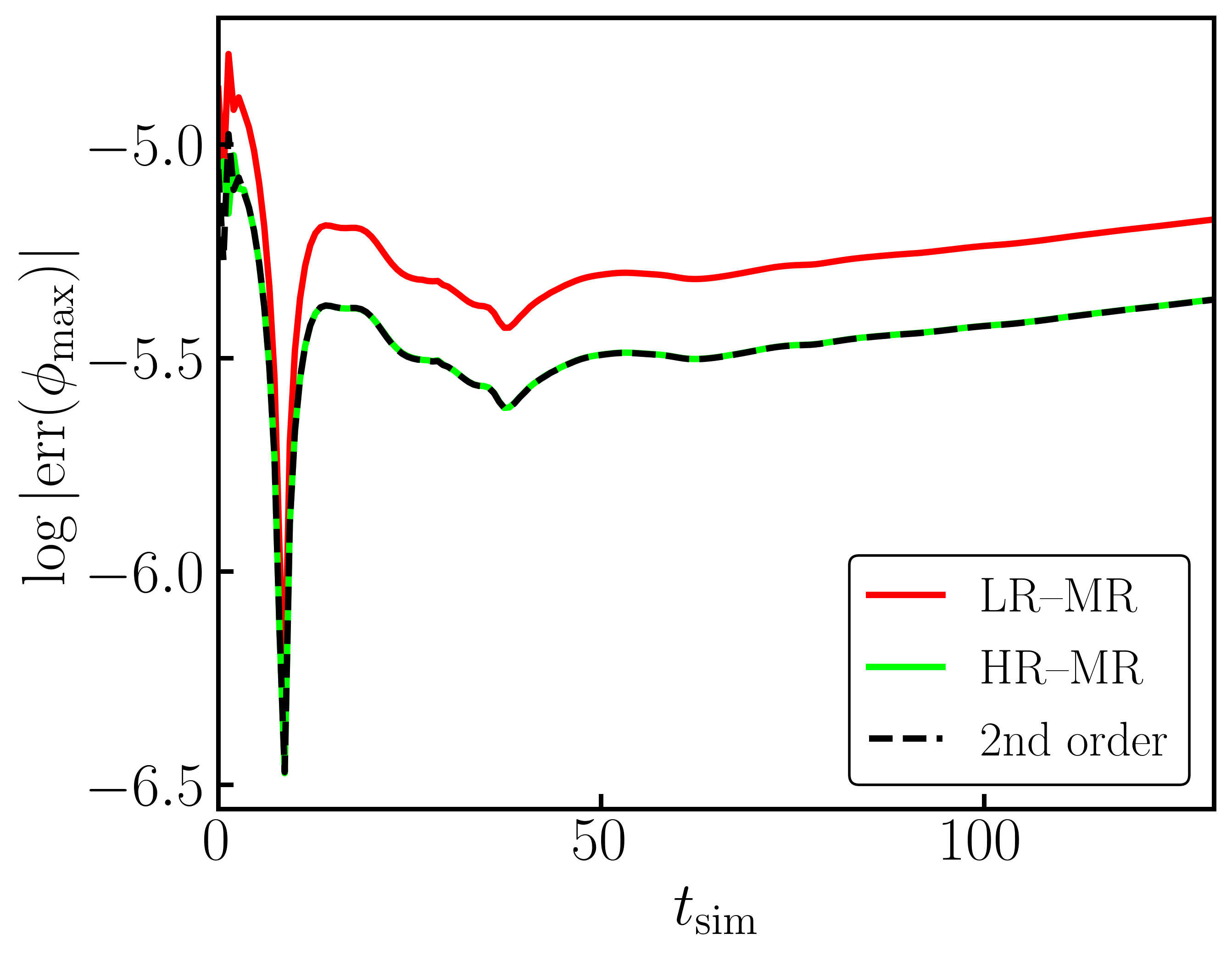}
\caption{Convergence test of the error in the evolution of the maximum of the scalar field $\mathrm{err}(\phi_\mathrm{max})$ between simulations with different resolutions, showing 2nd order convergence. The labels refer to the differences between low-resolution (LR), medium-resolution (MR) and high-resolution (HR).}
\label{fig:convergence}
\end{figure}

To convergence test our results, in Fig. \ref{fig:convergence} we compare the evolution of the maximum of the scalar field for simulations with three resolution $N^3=\lbrace 192^3, 224^3, 256^3\rbrace$. We find that the error decreases consistent with 2nd order convergence. 

\begin{figure}[b!]
\centering
\includegraphics[width=\columnwidth]{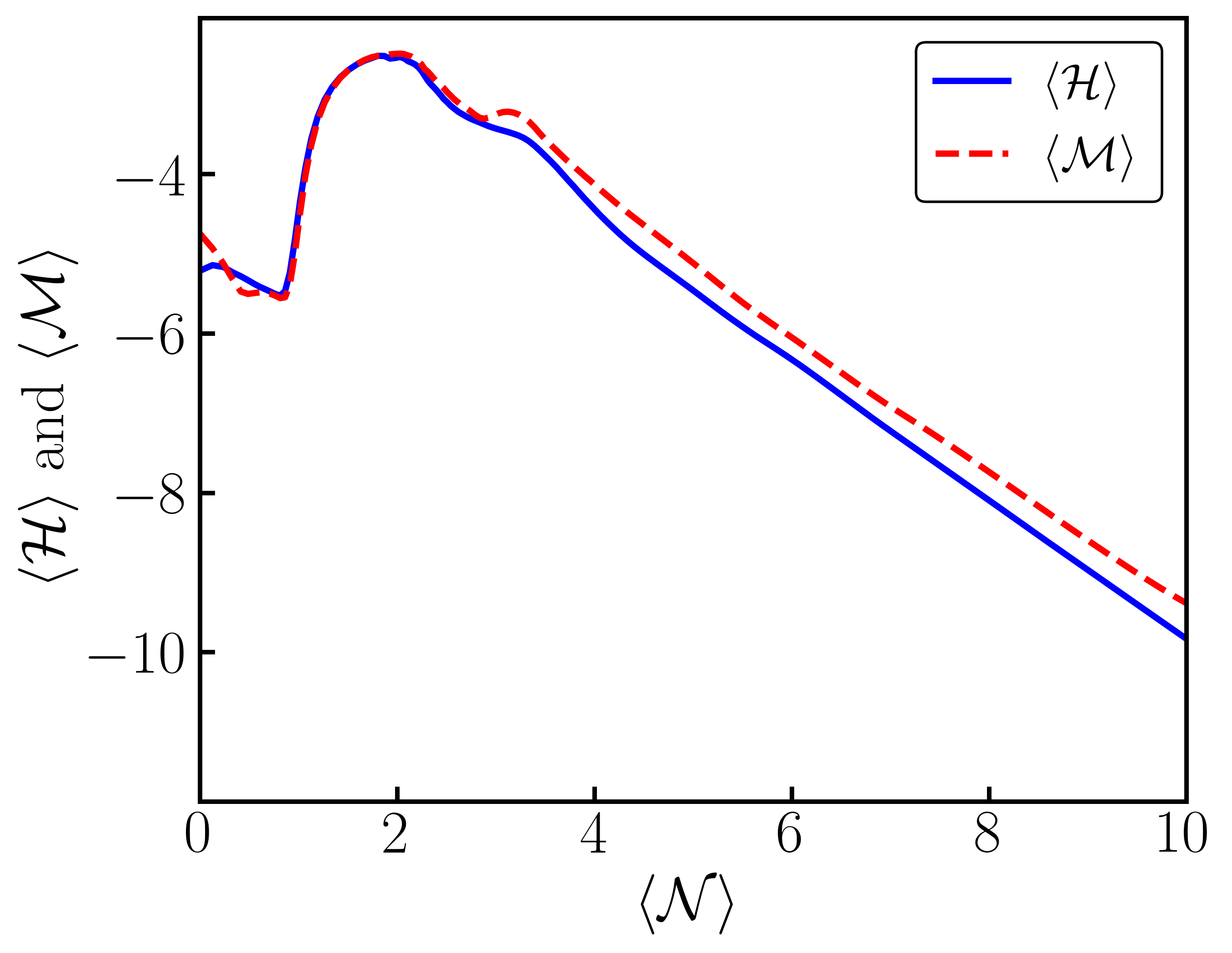}
\caption{Evolution of the averaged Hamiltonian and momentum constraints for the simulation with super-Planckian characteristic scale, and of largest gradient and kinetic energy.}
\label{fig:Ham_Mom}
\end{figure}

We also monitor the evolution of the Hamiltonian and momentum constraints, shown in Fig. \ref{fig:Ham_Mom}. At $\langle N \rangle \approx 1$ we observe a sharp peak in both constraints due to black hole formation, which is inflated away shortly after.

\clearpage

\end{document}